\documentclass[9pt,twocolumn,twoside]{opticajnl}
\journal{opticajournal} % use for journal or Optica Open submissions

\setboolean{shortarticle}{true}
\usepackage{lineno}
\title{Erbium-doped lithium niobate waveguide amplifier enhanced by an inverse-designed on-chip reflector}

\author[1]{Zhiwei Wei}
\author[1]{Jiangwei Wu}
\author[1]{Chengyu Chen}
\author[1]{Hao Li}
\author[1,3]{Wenjie Wan}
\author[1,2]{Xianfeng Chen}
\author[1,*]{Yuping Chen}

\affil[1]{State Key Laboratory of Photonics and Communications, School of Physics and Astronomy, Shanghai Jiao Tong University, Shanghai 200240, China}
\affil[2]{Collaborative Innovation Center of Light Manipulations and Applications, Shandong Normal University, Jinan 250358, China}
\affil[3]{University of MichiganShanghai Jiao Tong University Joint Institute, Shanghai Jiao Tong University, Shanghai, China}

\affil[*]{ypchen@sjtu.edu.cn}

\begin{abstract}
This study presents a 3.6-cm-long erbium-doped lithium niobate waveguide amplifier enhanced by an inverse-designed on-chip reflector. Integrating the reflector at the waveguide end yielded an internal net gain of 40.5 dB, achieving a 17.3 dB gain improvement compared to a comparable reflector-free amplifier under small signal conditions. By eliminating bidirectional pumping requirements, the system complexity was reduced. These results highlight a novel strategy for optimizing integrated optical amplifiers, combining high gain with simplified architecture. The approach holds promise for advancing high-density photonic integrated systems, demonstrating the efficacy of inverse design in tailoring photonic device performance for practical applications.
\end{abstract}

\setboolean{displaycopyright}{false} % Do not include copyright or licensing information in submission.

\begin{document}

\maketitle
%Example with the corresponding author designated by an asterisk:

%\author{Author One\authormark{1} and Author Two\authormark{2,*}}

%\address{\authormark{1}Peer Review, Publications Department,
%Optica Publishing Group, 2010 Massachusetts Avenue NW,
%Washington, DC 20036, USA\\
%\authormark{2}Publications Department, Optica Publishing Group,
%2010 Massachusetts Avenue NW, Washington, DC 20036, USA\\
%%\authormark{3}xyz@optica.org}

%\email{\authormark{*}xyz@optica.org}}

%Example with the corresponding author designated by an asterisk and a note indicating equal contributions by two authors.

%\author{Author One\authormark{1,3} and Author %Two\authormark{2,3,*}}

%\address{\authormark{1}Peer Review, Publications Department,
%Optica Publishing Group, 2010 Massachusetts Avenue NW, %Washington, DC 20036, USA\\
%\authormark{2}Publications Department, Optica Publishing Group, %2010 Massachusetts Avenue NW, Washington, DC 20036, USA\\
%\authormark{3}The authors contributed equally to this work.\\
%\authormark{*}xyz@optica.org}}

%\section{Examples of Article Components}
%\label{sec:examples}
As the cornerstone of modern telecommunications infrastructure, the erbium-doped fiber amplifier (EDFA) has been essential for high-capacity, long-haul data transmission since its invention in the late 1980s\cite{ref1}. The unique optical properties of erbium ions, particularly their stable transitions and long excited-state lifetime, enable broad gain spectra across the conventional telecommunications C-band. This makes EDFAs particularly well-suited for amplifying multiple wavelengths at high bit rates. The success of erbium-doped fibers has driven efforts to integrate erbium ions into other waveguide platforms, leading to the development of erbium-doped waveguide amplifiers (EDWAs)\cite{ref2}. Over the years, various host materials and waveguide designs have been explored, including ion-exchanged glass and lithium niobate waveguides, as well as planar waveguides in materials like $Al_2O_3$, $Si_3N_4$, and $Ta_2O_5$\cite{ref3,ref4,ref5,ref6,ref7}. Among these materials, thin-film lithium niobate (TFLN) has become a preferred material in integrated optics due to its exceptional optical characteristics, including a high refractive index, broad transparency range, and strong electro-optic, acoustic-optic, and nonlinear optical responses\cite{ref8,refshi}. This versatile platform has enabled numerous applications, such as ultra-compact frequency combs\cite{ref9,ref10}, high-speed electro-optic modulators\cite{ref11,ref12}, efficient wavelength converters\cite{ref13,ref14}, advanced sensors\cite{refwang1,refwang2,refyuanfano}, optical phased arrays\cite{refwuOPA}, and integrated lasers\cite{ref15,ref16,refwu,refbo}. Erbium-doped lithium niobate waveguide amplifiers have recently received significant attention from researchers\cite{refw1,refw2,refw3,refw4,refw5,refw6,refw7,refw8,refw9,refchen,refchen2}. As the pump light propagates further along the waveguide amplifiers, the intensity of the pump light decreases due to losses, which reduces the amplification effect of the signal light. Bidirectional pumping can improve this issue, resulting in a high pump intensity distribution at both ends and low in the middle. Compared to forward and backward pumping, bidirectional pumping has been demonstrated by researchers to provide the best amplification effect. It can improve the gain by about 1 dB under higher pump power conditions\cite{refw5}. Here a new principle was proposed to improve the gain. An inverse-designed on-chip reflector was introduced into the end of the erbium-doped lithium niobate waveguide amplifier that can improve the internal net gain by about 17.3 dB compared to a comparable reflector-free amplifier.

\begin{figure*}[ht]
\flushleft
\centering
\includegraphics[]{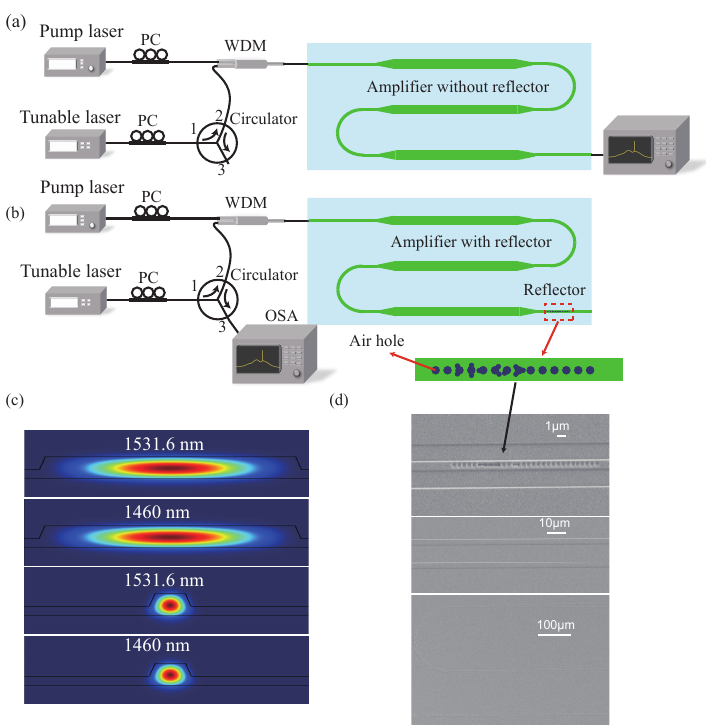}
\caption{Schematic illustration of the experimental setup. PC: polarization controller, WDM: wavelength division multiplexer, OSA: optical spectrum analyzer. (a) Schematic illustration of the experimental setup without a reflector. (b) Schematic illustration of the experimental setup with a reflector. (c) Simulated fundamental TE mode in 9-$\mu$m-wide 
 and 1.1-$\mu$m-wide waveguide at 1460 nm and 1531.6 nm. (d) Scanning electron microscope (SEM) image of the waveguide and reflector.}\label{figure1}
\end{figure*}

The schematic illustration of the amplifier without a reflector and with a reflector is shown in Fig.\ref{figure1}(a) and Fig.\ref{figure1}(b).  Light from the fiber was coupled into a waveguide by an edge coupler and then went through an adiabatic waveguide taper until the width of the waveguide was 1.1 $\mu$m. Then, it went through an adiabatic waveguide taper until the width of the waveguide was 9 $\mu$m. In the bend region, to reduce the bend loss, the width of the waveguide was set as 2 $\mu$m, and the bend radius was set as 200 $\mu$m. At the end of the amplifier, the reflector was introduced in the waveguide with a width of 1.1 $\mu$m.  The simulated fundamental TE mode in 9-$\mu$m-wide and 1.1-$\mu$m-wide waveguide at 1460 nm and 1531.6 nm is shown in Fig.\ref{figure1}(c). The loss of the edge coupler in our experiment was about -7.7 dB/facet. The scanning electron microscope (SEM) image of the waveguide and reflector are shown in Fig.\ref{figure1}(d). 

The design of the reflector consists of two steps. First, similar to a Bragg reflector, a reflector for 1531.6 nm was designed which achieves light reflection through periodic air holes. The period of the air hole is 792.1 nm and the top radius of the air hole is 263 nm. Inverse design enables researchers to systematically determine the parameters of photonic devices based on predefined objectives \cite{RN37, RN38, RN39, RN40, RN41}. The workflow of inverse design involves four key steps:

\begin{enumerate}
    \item \textbf{Initialization}: Define a device or model composed of elements with tunable parameters \( x \).
    \item \textbf{Objective Function}: Establish a measurable goal \( f(x) \), such as maximizing transmission or minimizing loss.
    \item \textbf{Optimization}: Employ computational algorithms to iteratively adjust \( x \), aiming to optimize \( f(x) \).
    \item \textbf{Output}: Obtain the optimized parameters that best satisfy the objective.
\end{enumerate}

 In this work, the inverse design was used to improve the reflector's reflectivity by introducing air holes with random positions and different sizes. The objective function was formulated as:

\begin{equation}
    f(x) = \max R
\end{equation}

\begin{figure}[ht]
\flushleft
\centering
\includegraphics[]{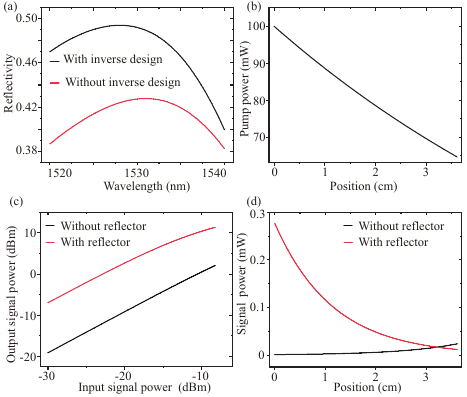}
\caption{Simulation result. (a) Comparison of the reflectivity between the reflector with inverse design(black) and the reflector without inverse design(red). (b) Relation between the on-chip pump power and position of the waveguide. (c)  Relation between the output signal power and input signal power. (d) Relation between the signal power and position of the waveguide.}\label{figure2}
\end{figure}

 R is the reflectivity of the signal light with a wavelength of 1531.6 nm. The sidewall of the lithium niobate waveguide has an angle, and the air holes should not be too small. The unit of the inverse design was an air hole with a radius of 222 nm. The algorithm used in the inverse design was a genetic algorithm, which is similar to our previous work\cite{refwei}. Considering a compact 10-$\mu$m-long reflector, a comparison of the reflectivity between the reflector with inverse design and the reflector without inverse design was shown in Fig.\ref{figure2}(a). The black line shows the result with inverse design and the red line shows the result without inverse design. The performance of reflectors with inverse design in the range of 1520 nm to 1540 nm is better than that of reflectors without inverse design. At 1531.6 nm, the reflectivity of the reflector with inverse design is 0.49 and the reflectivity of the reflector with inverse design is 0.43. The inverse design improved the reflectivity by 14\%. Increasing the length of the reflector can improve the reflectivity, while the extent to which the inverse design can enhance it will decrease accordingly.

Here a quasi-two-level model\cite{refmodel} was used in our simulation. Ignoring the effects of spontaneous emission and excited-state absorption, the rate equations can be written as:
\begin{equation}  
\frac{dN_1}{dt} = -\left(R_{12} + W_{12}\right) N_1 + \left(A_{21} + R_{21} + W_{21}\right) N_2  
\end{equation}  

\begin{equation}  
\frac{dN_2}{dt} = -\frac{dN_1}{dt}  
\end{equation}

\begin{equation}  
N_0 = N_1 + N_2  
\end{equation} 
 $N_1$ is the population density of the energy level 1 ($^4I_{15/2}$), and $N_2$ is the population density of the energy level 2 ($^4I_{13/2}$). $N_0$ is the total erbium-dopant concentration. The doping concentration of erbium-doped lithium niobate is 1 mol\%, which corresponds to an $Er^{3+}$ ion concentration of \(1.9 \times 10^{20} \, \text{cm}^{-3}\). The absorption and emission rates for the pump (signal) are denoted as \( R_{12} \) \( W_{12} \)  and \( R_{21} \) \( W_{21} \), respectively. The spontaneous transitions from the excited state to the ground state are represented by the rate \( A_{21} = \frac{1}{\tau} \), where \( \tau \) is the fluorescence lifetime.  

In the case of steady state, the pump evolution and signal evolution along the propagation direction can be written as:

\begin{equation}  
\frac{d i_p}{d z} = \left[-\alpha_p + \sigma_{p,21} N_2 - \sigma_{p,12} N_1\right] i_p  
\end{equation}  

\begin{equation}  
\frac{d i_s}{d z} = \left[-\alpha_s + \sigma_{s,21} N_2 - \sigma_{s,12} N_1\right] i_s  
\end{equation}  
 $i_p$ is the intensity of the pump light, and $i_s$ is the intensity of the signal light. $\alpha_p$ is the propagation loss of pump light, and $\alpha_s$ is the propagation loss of signal light. $\sigma_{p,21}$ is the absorption cross-section of the pump light, and $\sigma_{p,12}$ is the emission cross-section of the pump light. $\sigma_{s,21}$ is the absorption cross-section of the signal light, and $\sigma_{s,12}$ is the emission cross-section of the signal light.

Considering a 3.6-cm-long waveguide, the pump evolution along the propagation direction is shown in Fig.\ref{figure2}b. As propagation distance increases, the pump power decreases. After being amplified through forward propagation, the signal light is reflected by introducing a reflector at the end of the amplifier. After reflection, the intensity of the pump light gradually increases during the propagation of the signal light. In this way, the amplification effect of the signal light can be significantly enhanced.  Considering the pump light power to be 100 milliwatts, the relation between the  output signal light power and the input signal power is shown in Fig.\ref{figure2}c. The black line shows the result without a reflector and the red line shows the result with a reflector. Considering the input signal light power as 1 microwatt and the pump light power as 100 milliwatts, the signal light power at different positions is shown in Fig.\ref{figure2}d. During the small signal conditions, the amplification effect with a reflector is more than 13 dB higher than without a reflector.

The schematic illustration of the experimental setup is shown in Fig.\ref{figure1}. Signal light propagates through polarization controllers (PC) and a circulator. Pump light propagates through polarization controllers (PC). Then they are multiplexed by a wavelength division multiplexer (WDM), and coupled to the erbium-doped TFLN chip with a thickness of 600 nm via a lensed fiber. The etching depth of the erbium-doped TFLN waveguide is 350 nm. In the amplifier with a reflector, the amplified signal light was collected by the same lensed fiber. Then the amplified signal light propagated through the WDM and circulator and was detected by the optical spectrum analyzer (OSA). In the amplifier without a reflector, the amplified signal light was collected by another lensed fiber. Then the amplified signal light was detected by the OSA. The signal light is generated by an external-cavity tunable laser operating in the telecom band (New Focus, 1520-1570 nm), with its power regulated using an optical attenuator. A 1460-nm laser
serves as the pump laser. The polarization states of both the signal and pump light are set to the $TE_{00}$ mode using PC to ensure minimal loss and maximum net gain. The device is fabricated on erbium-doped thin film lithium niobate by standard electron beam lithography and reactive ion etching. Details of the fabrication process can be found in our previous work\cite{RN36}.

\begin{figure}[ht]
\flushleft
\centering
\includegraphics[]{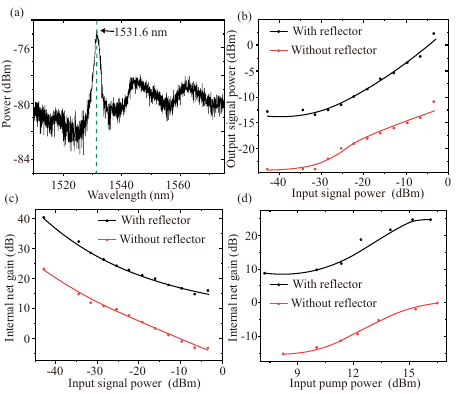}
\caption{(a) Comparison of the internal net gain between the amplifier with reflector and the amplifier without reflector. (b) Comparison of the on-chip amplified power between the amplifier with reflector and the amplifier without reflector.   (c) Relation of the amplifier with a reflector between the on-chip pump power and the on-chip internal net gain. (d) Relation of the amplifier without a reflector between the on-chip pump power and the on-chip internal net gain. }\label{figure4}
\end{figure}

The performance characterization of the amplifier is shown in Fig.\ref{figure4}. The amplified spontaneous emission (ASE) spectrum of the EDWA is shown in Fig.\ref{figure4}(a). The maximum gain marked by the green dashed line was at 1531.6 nm, which is also the selected wavelength for the gain characterization. The calculated propagation loss at 1531.6 nm is -1.44 dB/cm, using the measured Fabry–Perot fringes\cite{refloss}. The comparison of the output signal power between the amplifier with a reflector and the amplifier without a reflector is shown in Fig.\ref{figure4}b. With an on-chip input pump power of 16.7 dBm and an on-chip input signal power of -42.9 dBm, the on-chip output signal power with a reflector is -12.8 dBm while the on-chip output signal power without a reflector is -23.9 dBm. Under small signal amplification conditions, the output signal power of the amplifier with a reflector is 11.1 dB (12.9 times) greater than that of the amplifier without a reflector. The internal net gain can be used to characterize the amplifier's performance, as shown below.

\begin{equation}
G = 10 \log_{10} \frac{P_{out}}{P_{in}} -\alpha L
\end{equation}
 G is the internal net gain. $P_{out}$ is the output signal power and $P_{in}$ is the input signal power. $\alpha$ is the propagation loss including  the scattering loss and the absorption loss\cite{refw6}. $L$ is the propagation distance. In the calculation of internal net gain, the propagation loss is approximated to be -1.44 dB/cm. A comparison of the internal net gain between the amplifier with a reflector and the amplifier without a reflector is shown in Fig.\ref{figure4}(c). The black line shows the result with a reflector and the red line shows the result without a reflector. As the input signal power increases, the internal net gain decreases. The amplification performance of the amplifier with a reflector is significantly higher than that of the amplifier without a reflector. Under small signal conditions, with a fixed pump power of 16.7 dBm, the internal net gain of the amplifier with a reflector can reach 40.5 dB, while the internal net gain without a reflector is 23.2 dB. Compared to the amplifier without a reflector, the internal net gain of the amplifier with a reflector is improved by 17.3 dB(a 53.7-fold increase).  With an on-chip input signal power of -23.8 dBm, the relation of the on-chip pump power and the on-chip internal net gain is shown in Fig.\ref{figure4}d. The black line shows the result with a reflector and the red line shows the result without a reflector. Under varying pump power conditions, the amplification effect of the signal light in the amplifier with a reflector remains significantly higher than that in the amplifier without a reflector.

In conclusion, this research successfully realized a high-performance erbium-doped lithium niobate waveguide amplifier using inverse design. The on-chip reflector not only dramatically improved signal amplification (achieving 40.5 dB internal net gain) but also simplified system architecture by enabling unidirectional pumping. Experimental results demonstrated that the amplifier incorporating an inverse-designed reflector achieved a 17.3 dB (a 53.7-fold increase) improvement in internal net gain and a 12.9-fold increase in output signal power compared to its counterpart without the reflector. Combining simulations and experiments, this work highlights the potential of inverse-designed structures in optimizing photonic devices, offering a valuable reference for developing high-gain, low-complexity integrated amplifiers. These advancements could accelerate progress in optical communications and on-chip photonic systems.

\begin{backmatter}
\bmsection{Funding} This work was supported by the National Natural Science Foundation of China (Grant No. 12134009, Grant No.12474335 )

\bmsection{Acknowledgments}  We acknowledge the Center for Advanced Electronic Materials and Devices (AEMD) of Shanghai Jiao Tong University for the fabrication support.

\bmsection{Disclosures} The authors declare no conflicts of interest.

\smallskip

\bmsection{Data Availability Statement} Data underlying the results presented in this paper are not publicly available at this time but may be obtained from the authors upon reasonable request.

\bigskip

\end{backmatter}
% Bibliography
\bibliography{sample}

% Full bibliography added automatically for Optics Letters submissions; the following line will simply be ignored if submitting to other journals.
% Note that this extra page will not count against page length
\bibliographyfullrefs{sample}

%Manual citation list
%\begin{thebibliography}{1}
%\bibitem{Zhang:14}
%Y.~Zhang, S.~Qiao, L.~Sun, Q.~W. Shi, W.~Huang, %L.~Li, and Z.~Yang,
 % \enquote{Photoinduced active terahertz metamaterials with nanostructured
  %vanadium dioxide film deposited by sol-gel method,} Opt. Express \textbf{22},
  %11070--11078 (2014).
%\end{thebibliography}

% Please include bios and photos of all authors for aop articles
\ifthenelse{\equal{\journalref}{aop}}{%
\section*{Author Biographies}
\begingroup
\setlength\intextsep{0pt}
\begin{minipage}[t][6.3cm][t]{1.0\textwidth} % Adjust height [6.3cm] as required for separation of bio photos.
  \begin{wrapfigure}{L}{0.25\textwidth}
    \includegraphics[width=0.25\textwidth]{john_smith.eps}
  \end{wrapfigure}
  \noindent
  {\bfseries John Smith} received his BSc (Mathematics) in 2000 from The University of Maryland. His research interests include lasers and optics.
\end{minipage}
\begin{minipage}{1.0\textwidth}
  \begin{wrapfigure}{L}{0.25\textwidth}
    \includegraphics[width=0.25\textwidth]{alice_smith.eps}
  \end{wrapfigure}
  \noindent
  {\bfseries Alice Smith} also received her BSc (Mathematics) in 2000 from The University of Maryland. Her research interests also include lasers and optics.
\end{minipage}
\endgroup
}{}

\end{document}